\documentstyle[12pt,epsfig]{article}
\newlength{\dinwidth}
\newlength{\dinmargin}
\begin{document}
\newcommand{\be}{\begin{equation}}
\newcommand{\ee}{\end{equation}}
\newcommand{\ba}{\begin{eqnarray}}
\newcommand{\ea}{\end{eqnarray}}
\newcommand{\as}{\bar{\alpha}_s}
\newcommand{\dy}{\Delta Y}
\newcommand{\tdm}[1]{\mbox{\boldmath $#1$}}  
\titlepage
\begin{flushright}
IPPP/01/21 \\
DCTP/01/42 \\
3 May 2001 \\
\end{flushright}
\begin{center}
\vspace*{2cm}
{\Large \bf Azimuthal decorrelation of forward and backward jets at the 
Tevatron}  

\vspace*{1cm}
J.\ Kwieci\'{n}ski$^{a,b}$, A.D.\ Martin$^{a}$,  L. \ Motyka$^{c,d}$
and J.\ Outhwaite$^{a}$ \\
\vspace*{0.5cm}
$^{a}$ Department of Physics and Institute for Particle Physics Phenomenology, University of      
Durham, Durham, DH1 3LE \\
$^{b}$ H.\ Niewodnicza\'{n}ski Institute of Nuclear Physics, Krak\'{o}w, Poland \\
$^{c}$ THEP, Department of Radiation Sciences, Uppsala University, Sweden   \\
$^{d}$ Institute of Physics, Jagellonian University, Krak\'{o}w, Poland \\
\end{center}

\vspace*{2cm}

\begin{abstract}
We analyse the azimuthal decorrelation of \lq Mueller-Navelet\rq\ dijets produced in the 
$p\bar p$ collisions at Tevatron energies using a BFKL framework which incorporates dominant 
subleading effects.  We show that these effects significantly reduce the decorrelation yet 
they are still insufficient to give satisfactory description of experimental data.  
However a good description of the data is obtained after incorporating within 
formalism the effective rapidity defined by Del Duca and Schmidt.  
\end{abstract}

\newpage


An ideal test of $\log(1/x)$ dynamics is the ``Mueller-Navelet'' process 
\cite{MN,WJS,ANDER} in hadron-hadron collisions with one backward jet separated from a forward jet 
by a large rapidity interval. The test has been under study for several years,
but has recently been sharpened by the considerable improvement in the
precision of the data \cite{TEVDATA}.  The relevant hard scale is specified by
the transverse momenta of the jets $(\tdm{k}_1, \tdm{k}_2)$.  An advantage of this process is that 
these momenta can be of comparable magnitude so the intervening DGLAP evolution is suppressed. 
The behaviour of the cross-section is therefore sensitive to the diffusion of transverse momenta
of the intermediate gluons, which is a property of leading $\log(1/x)$ dynamics 
\cite{BFKL,LIPATOV}.  The characteristic features of this process are (i) a rapid increase of the 
2-jet production cross section with increasing incident energy and (ii) a
weakening of the azimuthal back-to-back correlation between the jets as the 
rapidity interval is increased. The analyses have so far been within the
leading order $\log(1/x)$ BFKL framework, and indicate much more decorrelation
than shown by the data.  The improvement of the 2-jet data has been accompanied by progress 
in understanding and quantifying the effect of sub-leading $\log(1/x)$ corrections.  
The full NLO BFKL ($\log(1/x)$) contributions have been obtained and found to be large
\cite{NLO1,ROSS,SALAM}.  However the major contribution can be resummed to all orders in a 
straightforward way \cite{LUNDCC,SUTTON1,STASTO}.  
To be precise the resummation is effected by requiring 
that the virtuality of the intermediate gluons is dominated by their transverse momenta squared, 
which although formally sub-leading, is implicit in the derivation of the BFKL equation.  We call 
this the consistency condition (CC). The resummation is found to stabilise the solution of the 
NLO BFKL equation.  The purpose of this paper is to incorporate those subleading effects 
generated by the consistency constraint in the theoretical description 
of the dijet production in hadronic collisions.

The cross-section describing the production of a forward jet $(x_1,k_1^2)$ and
a backward jet $(x_2,k_2^2)$, with relative azimuthal angle $\phi-\pi$, 
may be written
\be
{d\sigma \over dx_1 dx_2 dk_1^2 dk_2^2 d\phi} = {1 \over x_1 x_2}
\left[x_1 f_{\rm eff}(x_1,k_1^2) \; x_2 f_{\rm eff}(x_2,k_2^2) \right]	
{d\hat{\sigma} \over dk_1^2 dk_2^2 d\phi},
\label{eq:a1}
\ee
where $f_{\rm eff} = g + {4 \over 9}(q+\bar{q})$ is the effective parton 
distribution of the incoming proton (antiproton). Note, that with this 
convention the back-to-back configuration of the jets in the transverse plain 
corresponds to $\phi=0$. The hard partonic 
cross section (defined here for the gluon-gluon scattering)  is of the form
\be
{d\hat{\sigma} \over dk_1^2 dk_2^2 d\phi} = 
{\as^2 \pi \over 4 (k_1 k_2)^3 }
\sum_{m=0}^{\infty} \cos (m\phi) I_m(\dy,\rho)
\label{eq:a2}
\ee
where $\as=3\alpha_s/\pi$,
\be
\dy = \log\left({x_1 x_2 s \over k_1 k_2}\right), \;\;\;
\rho = \log\left({k_1^2 \over k_2^2}\right), 
\label{eq:a3}
\ee
and $I_m$ is the solution of the BFKL equation with the consistency condition imposed. In 
this paper we shall use a fixed strong coupling constant corresponding to $\as = 0.15$.  The 
solution may be written
\be
I_m(\dy,\rho) = \int_0^{\infty} d\nu \int {d\omega\over 2\pi i} \;\;
{1 \over \omega - \chi_m (\omega,\nu) }
e^{\omega\dy} \cos(\rho\nu)
\label{eq:a4}
\ee
where the kernels $\chi_m$ are given by\footnote{An alternative resummation of subleading 
$\log(1/x)$ effects, based on keeping the collinearly enhanced contributions \cite{FLOR}
is found to give the same result if $\psi(z)$ of (\ref{eq:a5}) is approximated by 
$1/z$, apart from the inclusion in \cite{FLOR} of running $\alpha_s$ and non-singular terms 
in the splitting function.}
\be
\chi_m(\omega,\nu) = \as
\left[ 2 \psi(1) - 
\psi \left( {m+\omega+1\over 2}+ i \nu \right) - 
\psi \left( {m+\omega+1\over 2}- i \nu \right) 
\right]
\label{eq:a5}
\ee
with $\psi$ as usual, the logarithmic derivative of the Euler Gamma function.

We evaluate the $d\omega$ integral in (\ref{eq:a4}) by
picking up the leading singularity $\omega_m^0(\nu)$ which is the 
solution of the implicit equation
\be
\omega_m^0(\nu) = \chi_m(\omega_m^0(\nu),\nu),
\label{eq:a6}
\ee
(cf.(\ref{eq:a5})), to obtain
\be
I_m(\dy,\rho) = \int_0^{\infty} 
d\nu R_m^0(\nu)\exp(\omega_m^0(\nu)\dy)\cos(\rho\nu)
\label{eq:a7}
\ee
where the residue $R$ is
\be
R_m^0(\nu) = \left[ 1 - 
{d\chi_m(\omega,\nu) \over d\omega} 
\left|_{\omega=\omega_m^0(\nu)} \right.
\right]^{-1}.
\label{eq:a8}
\ee
Hence at LO the above solution (\ref{eq:a6}) would simplify to
\be
\omega_m^0(\nu) = \chi_m(0,\nu),
\ee
with $R_m^0 = 1$.

\begin{figure}[h]
\begin{center}
\mbox{\epsfig{figure=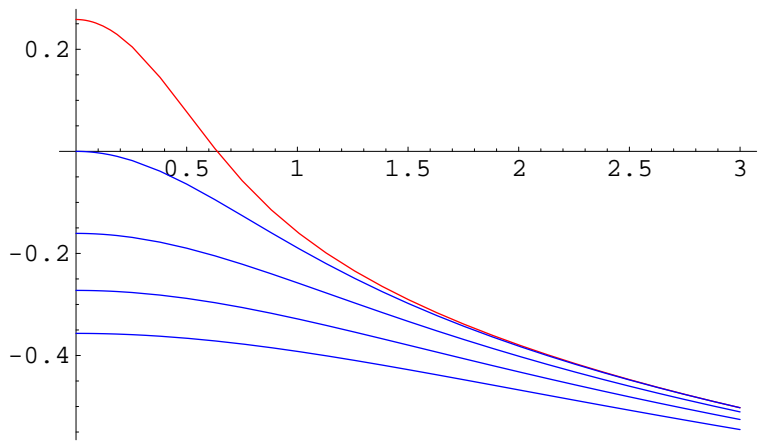,width=10cm}}
\caption{$\omega_m^0(\nu)$, for $0\le m \le 4$.  In decreasing order, the curves correspond to 
$m = 0, 1, \ldots 4$.}
\label{eqn:fig1}
\end{center}
\end{figure}

In Fig. \ref{eqn:fig1} we illustrate the solution $\omega_m^0(\nu)$, (\ref{eq:a6}), 
for various values of $m$.  For $\dy \gg 1$ the dominant contribution comes from the region
$\nu \sim 0$. If we then expand around the saddle point, $\nu = 0$, we see that 
\be
I_m(\dy,\rho) \sim \exp(\omega_m^0(0)\dy).
\label{eq:a9}
\ee

The total forward+backward jet cross section is controlled by the $m=0$ contribution and the 
intercept $\omega_0^0(0)$ gives the rate of growth as 
the rapidity interval increases. We emphasize that the ``subleading'' intercept
$\omega_0^0(0)$ is significantly smaller than its LO counterpart, and
significantly above the NLO approximation.

\begin{figure}
\leavevmode 
\begin{center} 
\epsfxsize = 12cm 
\epsfysize = 10cm 
{\large\bf a)}\hspace{1cm}
\epsfbox{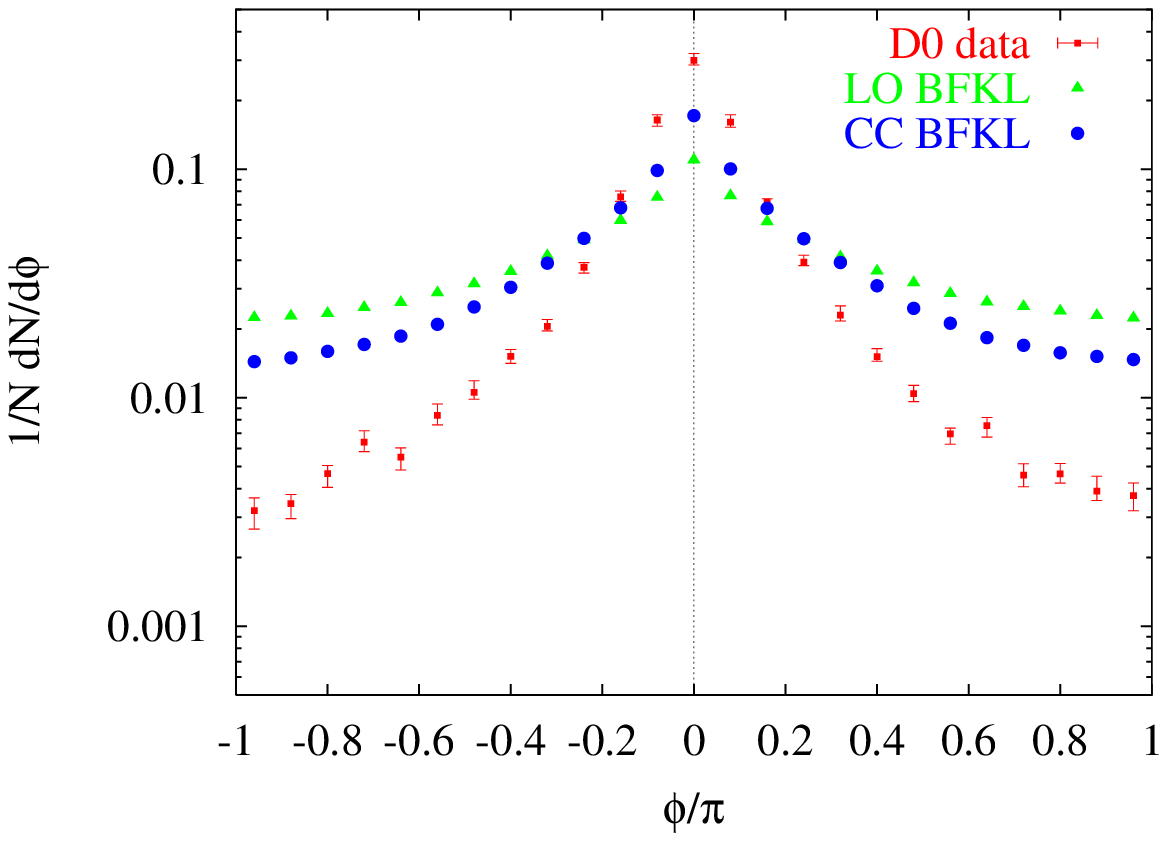}\\
\epsfxsize = 12cm 
\epsfysize = 10cm 
{\large\bf b)}\hspace{1cm}
\epsfbox{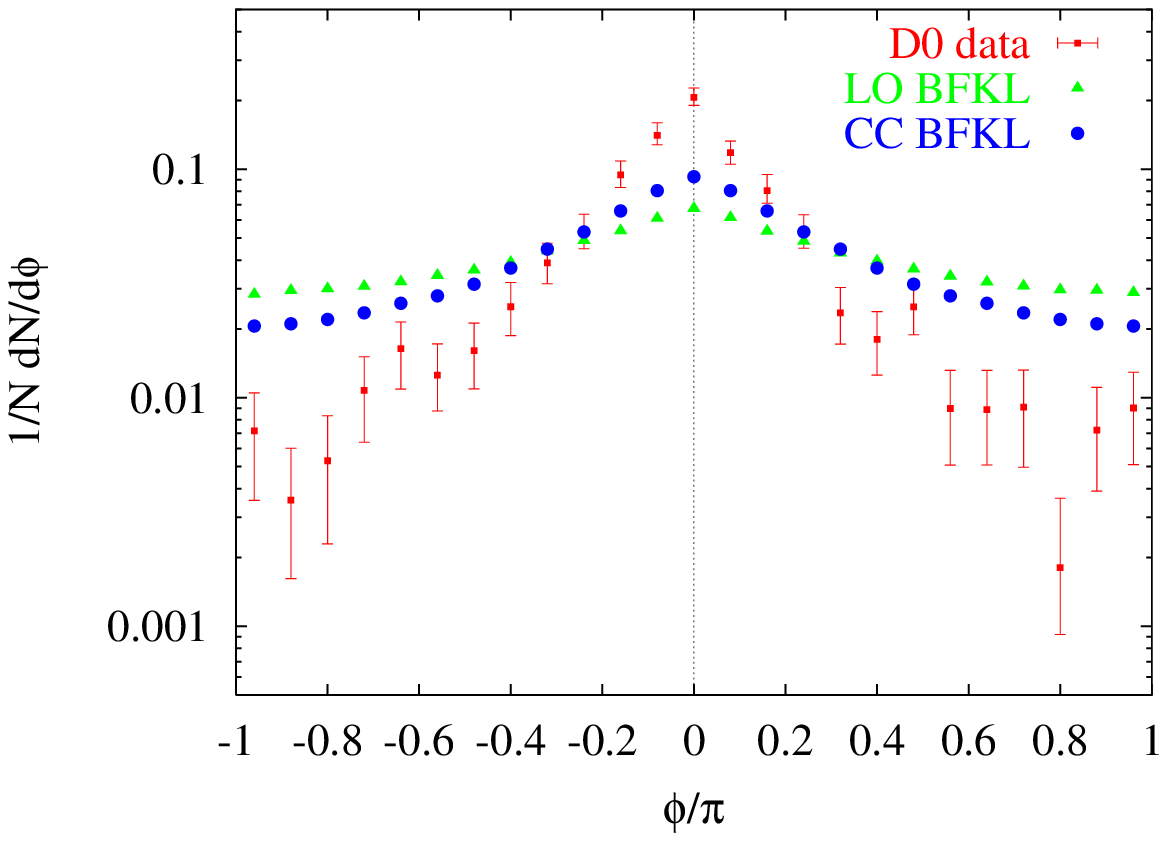}\\
\end{center} 
\caption
{
Azimuthal decorrelation of the forward and backward jets for two rapidity separation intervals:  
(a) $2.5 < \Delta\eta < 3.5$, (b) $4.5 < \Delta\eta < 5.5$.  The D0 data 
\cite{TEVDATA}  are shown by points with errorbars.  
The BFKL (LO) and BFKL (CC) predictions are shown by 
triangles and circles respectively.} 
\end{figure}

It can also be seen from Fig.~\ref{eqn:fig1} that the intercepts
$\omega_m^0(0)$ decrease with increasing $m$. This implies the
dominance of the $m=0$ term at large values of $\dy$, which
determines the rate of decorrelation of the azimuthal angle between
the forward and backward jets.  In practice we do not restrict ourselves to the saddle point 
approximation, but rather evaluate (\ref{eq:a7}) numerically.  In Fig.~2 we plot the 
azimuthal distributions $d\sigma/d\phi$
\be
{d\sigma\over d\phi} =\int_{\Delta} dx_1dx_2dk_1^2dk_2^2 
{d\sigma \over dx_1dx_2dk_1^2dk_2^2}
\label{eq:a10}
\ee
where the integration region $\Delta$ is restricted by the experimental
cuts~\cite{TEVDATA}.  To be precise the cuts which we impose are as follows: 
\begin{enumerate}
\item[(i)] the tagged jet tranverse momenta: $|\tdm{k}_i| > 20$~GeV, and 
$\max (|\tdm{k}_1|,  |\tdm{k}_2|) > 50$~GeV;
\item[(ii)] the tagged jet pseudorapidities $|\eta_i| < 3.0$ and the pseudorapidity distance 
between jets $|\Delta\eta = |\eta_2 - \eta_1|$ was required to fall into one of two windows: 
$2.5 < \Delta\eta < 3.5$ or $4.5 < \Delta\eta < 5.5$.
\end{enumerate}         
Fig.~2 shows the results corresponding both to the LO BFKL 
equation and those which are obtained from the BFKL equation with the 
subleading  effects 
generated by the consistency constraint. We confront our predictions with 
the experimental data obtained at the Tevatron.  We see from these figures 
that the subleading effects reduce azimuthal decorrelation obtained from 
the LO BFKL equation.  The resulting $\phi$ distributions are however still 
too flat when compared with the experimental data.

It has been pointed out in ref.\ \cite{DDS} that the approximations 
which are adopted within the BFKL formalism may not be accurate, particularly 
in the region away from the back-to-back configuration of the two jets.  
It has been proposed that in order to improve accuracy of the predictions 
one should use the effective rapidity difference $\hat y(m)$ for each azimuthal
projection $m$ in the place of $\Delta Y$ in the solutions $I_m$ of the BFKL 
equation(s). The effective  rapidity intervals $\hat y(m)$ are defined as:
\be
\hat y(m)=\Delta Y{\int d\phi \cos(m\phi)(d\sigma/dy_1dy_2dk_1dk_2d\phi)\
\over  \int d\phi \cos(m\phi)(d\sigma_0/dy_1dy_2dk_1dk_2d\phi)}
\label{eq:a11}
\ee
where $d\sigma/dy_1dy_2dk_1dk_2d\phi$ and $d\sigma_0/dy_1dy_2dk_1dk_2d\phi$ 
denote the exact $O (\alpha_s^3)$ contribution to the 
three jet cross section and its BFKL (i.e. large $\Delta Y$) approximation 
respectively.

Guided by this proposal we have used as the asymptotic variable the 
effective  rapidity $\hat y(\Delta Y,k_1^2,k_2^2,\phi)$
\be
\hat y(\Delta Y, k_1^2,k_2^2,\phi)=
\Delta Y {d\sigma/(dy_1dy_2dk_1dk_2d\phi)\
\over d\sigma_0/(dy_1dy_2dk_1dk_2d\phi)}
\label{eq:a12}
\ee
where $d\sigma_0/(dy_1dy_2dk_1dk_2d\phi)$ should now contain the subleading BFKL effects.
They just correspond to the replacement:
\be
\Delta Y \rightarrow \Delta Y - {1\over 2} 
\left| \log \left({k_1^2\over k_2^2}\right) \right| 
\label{eq:a13}
\ee
in the LO BFKL expression for $d\sigma_0/(dy_1dy_2dk_1dk_2d\phi)$.    

The use of effective rapidity $\hat y(\Delta Y, k_1^2,k_2^2,\phi)$  defined by equation 
(\ref{eq:a12}) aims at correcting  the large $\Delta Y$ approximation 
of the $2 \rightarrow 3$ jets  cross-section integrated over the rapidity of 
the third jet. In this approximation the integrated $2 \rightarrow 3$ jets 
cross-section is just proportional to $\Delta Y$.  
This is the result of the fact  that the 
rapidity $y_3$ of the third jet varies between the rapidities $y_1$ and $y_2$ 
of the tagged jets ($\Delta Y=y_2-y_1$)  and that, 
in the large $\Delta Y$ approximation,   
potential dependence of the momenta of the colliding partons 
on $y_3$ is neglected.   The analysis performed in ref.\ \cite{DDS} 
shows that this is an inaccurate approximation and that the use of the exact kinematics 
introduces a large suppression.  The effective rapidity 
$\hat y(\Delta Y, k_1^2,k_2^2,\phi)$  
incorporates this suppression and 
reflects the effective rapidity range spanned by the (mini)jets 
radiated with rapidities  between $y_1$ and $y_2$  \cite{DDS}.                 
In the formulae defining the exact 3-jet cross section a fixed
coupling was used corresponding to the value 
of $\as$ which entered the BFKL kernels.   
Let us mention, that there is a significant dependence of  
$\hat y(\Delta Y,k_1^2,k_2^2,\phi)$ on the $\phi$ variable. 
The conditional maximum with other variables fixed is always reached for 
$\phi = 0$ and if $k_1^2 \simeq k_2^2$ then 
$\hat y(\Delta Y,k_1^2,k_2^2,\phi=0) \simeq \Delta Y$.  
We have noticed that the numerical 
convergence of the integral in (\ref{eq:a7}) is poor if $\hat y$ is below 1.2.  
In this region we have therefore used the perturbative expansion of the solution 
of the (modified) BFKL equation up to $O(\alpha_s^3)$.  This required a careful 
treatment of the  3 jet cross section for the third jet with low transverse momentum $k_{3}$.  
Integration over the third jet phase space leads to infrared divergence which is cancelled
by the virtual correction to the Born term, when integrated over the region of phase space 
containing the singular point $k_{3}=0$.
 
\begin{figure}
\leavevmode 
\begin{center} 
\epsfxsize = 12cm 
\epsfysize = 10cm 
{\large\bf a)} \hspace{1cm}
\epsfbox{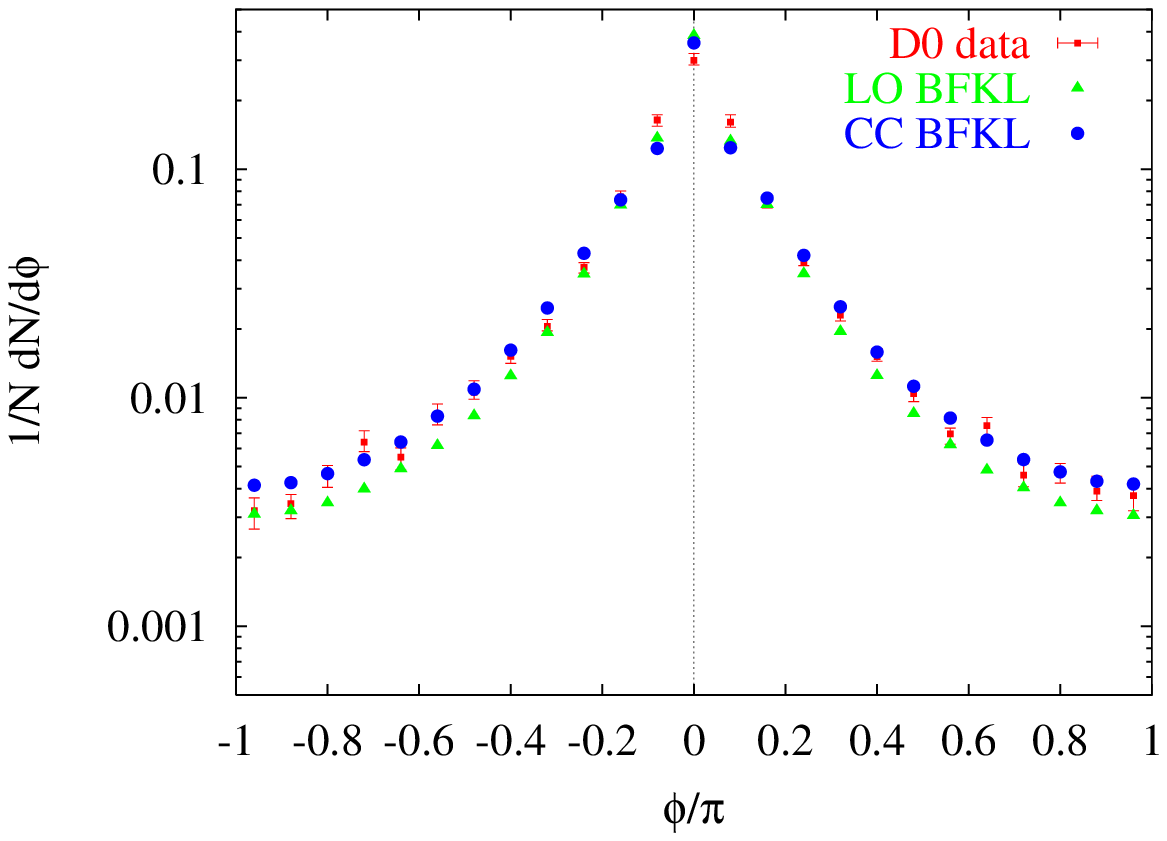} \\ 
\epsfxsize = 12cm 
\epsfysize = 10cm 
{\large\bf b)} \hspace{1cm}
\epsfbox{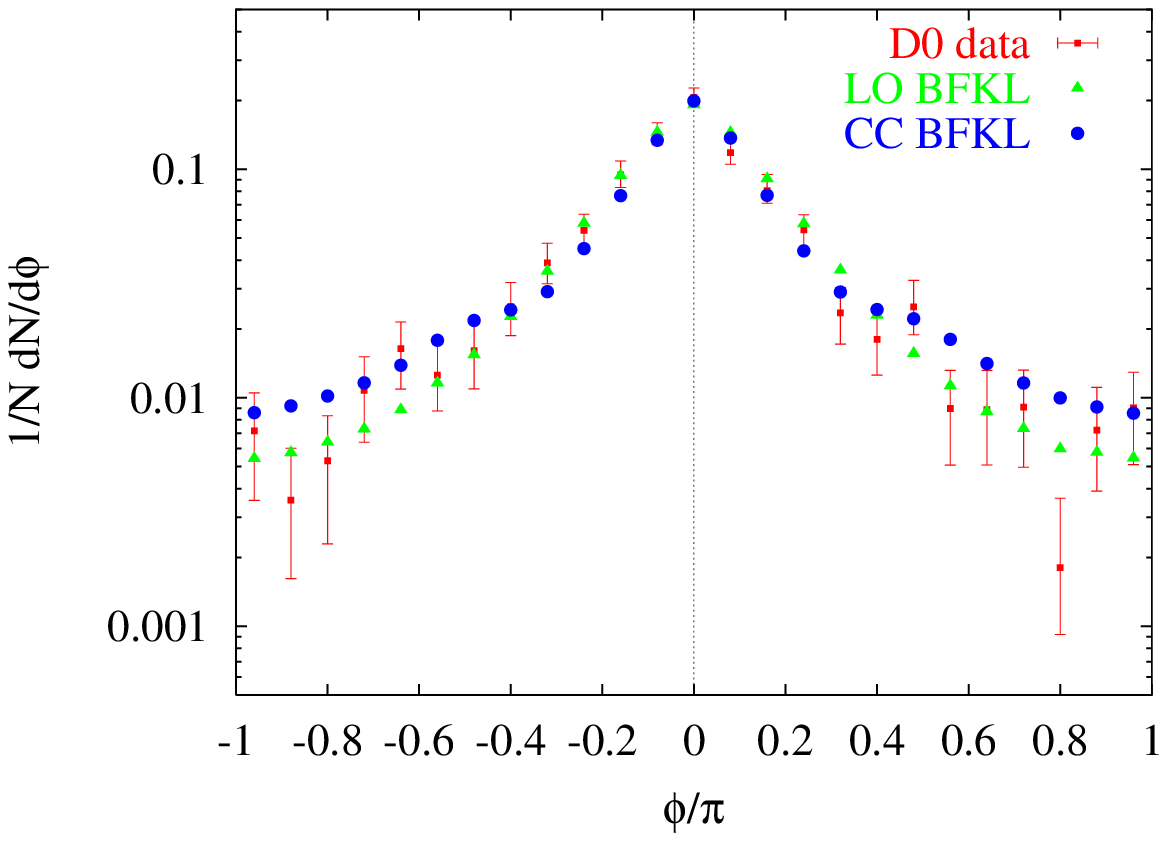} \\ 
\end{center} 
\caption{As for Fig.~2 but with the effective rapidity $\hat{y}$ of (\ref{eq:a12}) incorporated 
in the BFKL predictions.} 
\end{figure}

The results of the calculations incorporating the effective rapidity are summarised in Fig.~3.  
These plots show the $\phi$ distributions which are obtained from the BFKL framework with the 
subleading effects taken into account and compare with effective rapidity 
$\hat y(\Delta Y,k_1^2,k_2^2,\phi)$ in the place of $\Delta Y$ as an 
asymptotic variable.  Again we compare with the available experimental data.     
We see that the use of the effective rapidity significantly improves agreement 
with the data. In Figures~3(a,b) we also show results obtained from the LO BFKL 
equation using the corresponding effective rapidity.  
In contrast to the results of the BFKL calculations presented in Figures~2~(a,b), 
when the effective rapidity is used the azimuthal decorrelation is far from
being saturated (which would correspond to the distribution flat in $\phi$) 
and it should be more sensitive to the particular form of the BFKL kernel.  
Thus, it is surprising that these LO BFKL predictions do not differ appreciably 
from those obtained using the BFKL formalism with subleading 
effects {\it and} the effective rapidity.  This implies that the implementation of 
effective rapidity suppresses the potentially large effect of subleading corrections. 
It is mainly caused by the fact that the absolute values of the effective rapidities 
at the tails of the distributions are rather small (i.e. $\hat y \sim 1$), where 
the cross-section is dominated by the 2 jet + 3 jet production. 
In fact, the difference which is visible at the tails of the distributions
obtained from the BFKL~LO and BFKL~CC equations does not come from the difference 
in the cross sections at the tails but 
rather from the region of central $\phi$ which influences the tails
through entering the common normalisation factor $N$. 
Finally, we have checked, that the sensitivity
of the results to variations of the fixed $\as$ between 0.13 and 0.18 is small
and $\as = 0.15$ gives the best description of the data.


With experimental cuts used in the D0 analysis the cross-section
for cental bins of $\phi$ contains contributions from configurations
with $\tdm k_1 + \tdm k_2 = 0$. In order to describe properly
these configurations the full BFKL resummation is necessary. 
This is due to the fact that the kinematical suppression of
additional gluon emissions is weak in this case,  
and the corresponding effective rapidity $\hat y$ is  
close to $\Delta Y$. Thus, $\alpha_s \hat y$ becomes large
and it is not enough to retain only the first terms of
the perturbative expansion. In other words, the 
$O(\alpha_s^3)$ cross-section, calculated as the sum of the LO 3~jet and 
NLO 2~jet cross-section, is expected to be subject to
a series of large $~(\alpha_s \hat y)^n$ corrections. 
In fact, we have checked explicitly that without the resummation, the data
in the central $\phi$ bins are not well reproduced.   


We also studied the distribution of $\langle \cos\phi \rangle$ in the bins
of pseudorapidity separation of the jets $\Delta\eta$. Here we used only the
method and parameters which gave the best description of the azimuthal angle 
distributions of the Mueller-Navelet jets.  
The results of our calculation are compared to the experimental 
data~\cite{TEVDATA} in Fig.~4. Both the LO~BFKL and CC~BFKL 
equations with the effective rapidity $\hat y$ give a resonable description of the
experimental data, when the statistical and the correlated systematical errors
of the data are taken into account. However, it is clear that  some small 
discrepancy between the theoretical and experimental values still remains.      

\begin{figure}
\leavevmode 
\begin{center} 
\epsfxsize = 11cm 
\epsfysize = 8cm 
\epsfbox{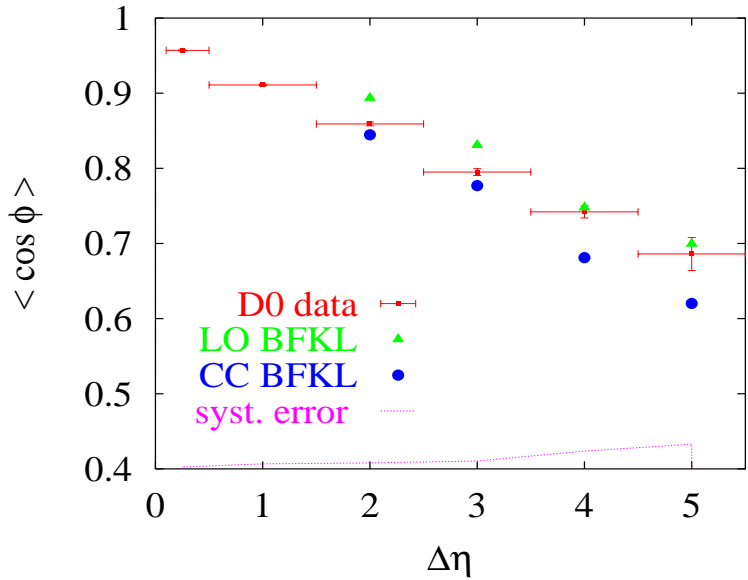} \\ 
\end{center} 
\caption{
The average value of $\cos\phi$ for various $\Delta\eta$ intervals.  The D0
data (points with errorbars) compared with the BFKL~LO calculations
(triangles) and with the BFKL~CC calculations (circles). The band between
the $\Delta\eta$ axis and the dotted line represents the
correlated systematic error of the data.
}  
\end{figure}

To summarise we emphasize the following results of our analysis:
\begin{enumerate}
\item[(i)] The BFKL formalism in its original form fails to describe the data on 
the azimuthal decorrelation of dijets.  The dominant subleading effects do improve 
the results obtained within the leading logarithmic approximation, but they still 
generate more decorrelation in the tails of the azimuthal distribution than that 
required by the data. 

\item[(ii)] Most of the relevant contribution to the decorrelation in the Tevatron
domain is present already in the fixed order $O(\alpha_s^3)$ expressions.  

\item[(iii)] In order to describe the data it is essential to use the 
effective rapidity $\hat{y}$ defined by eq.~(\ref{eq:a12}) and to use the 2 jet (LO + NLO) + 3 
jet cross-section in the domain of low values of $\hat y$.  
In this region the BFKL formalism does not therefore introduce any new information, which 
would not be present in the fixed order calculation.       

\item[(iv)] The only places where the use of the complete BFKL framework seems to be 
necessary are in the very central bins of $\phi$. 

\item[(v)] We therefore conclude that it is possible to describe the decorrelation data 
using the fixed order perturbative expressions in the tails of the azimuthal distribution 
and the complete BFKL resummation for $\phi \approx 0$. 
\end{enumerate}

\section*{Acknowledgements}

We thank Vittorio Del Duca for providing us with a copy of his programme to calculate the effective  rapidity.  
We thank the UK Particle Physics and Astronomy Research Council and the EU Framework TMR  
programme, contract FMRX-CT98-0194 (DG 12-MIHT) for support. 
LM is grateful to the Swedish Natural Science Research Council 
for the postdoctoral fellowship. This research has also been supported 
in part by the Polish State Committee for Scientific Research grants  
5 P03B 051 19 and 5 P03B 144 20.  
\\

\end{document}